\documentclass[modern]{aastex61}

\submitjournal{ApJ}

\shorttitle{Universal spin-mass relation}
\shortauthors{Scholz et al.}

\begin{document}

\title{A universal spin-mass relation for brown dwarfs and planets}

\correspondingauthor{Aleks Scholz}
\email{as110@st-andrews.ac.uk}

\author[0000-0001-8993-5053]{Aleks Scholz}
\affil{SUPA, School of Physics \& Astronomy, University of St Andrews, North Haugh, St Andrews, KY16 9SS, United Kingdom}

\author{Keavin Moore}
\affiliation{Faculty of Science, York University, 4700 Keele Street, Toronto, ON M3J 1P3, Canada}

\author{Ray Jayawardhana}
\affiliation{Faculty of Science, York University, 4700 Keele Street, Toronto, ON M3J 1P3, Canada}

\author{Suzanne Aigrain}
\affiliation{Subdepartment of Astrophysics, University of Oxford, Oxford, OX1 3RH, United Kingdom}

\author{Dawn Peterson}
\affiliation{Space Science Institute, 4750 Walnut Street, Suite 205, Boulder, CO 80301, USA}

\author{Beate Stelzer}
\affiliation{Institut f{\"u}r Astronomie und Astrophysik, Eberhard Karls Universit{\"a}t, Sand 1, D-72076 T{\"u}bingen, Germany}

\begin{abstract}
While brown dwarfs show similarities with stars in their early life, their spin evolution is much more akin to that of planets. We have used lightcurves from the K2 mission to measure new rotation periods for 18 young brown dwarfs in the Taurus star-forming region. Our sample spans masses from 0.02 to 0.08$\,M_{\odot}$ and has been characterised extensively in the past. To search for periods, we utilize three different methods (autocorrelation, periodogram, Gaussian Processes). The median period for brown dwarfs with disks is twice as long as for those without (3.1 vs. 1.6\,d), a signature of rotational braking by the disk, albeit with small numbers. With an overall median period of 1.9\,d, brown dwarfs in Taurus rotate slower than their counterparts in somewhat older (3-10\,Myr) star-forming regions, consistent with spin-up of the latter due to contraction and angular momentum conservation, a clear sign that disk braking overall is inefficient and/or temporary in this mass domain. We confirm the presence of a linear increase of the typical rotation period as a function of mass in the sub-stellar regime. The rotational velocities, when calculated forward to the age of the solar system assuming angular momentum conservation, fit the known spin-mass relation for solar system planets and extra-solar planetary-mass objects. This spin-mass trend holds over six orders of magnitude in mass, including objects from several different formation paths. Our result implies that brown dwarfs by and large retain their primordial angular momentum through the first few Myr of their evolution.   
\end{abstract}


\section{Introduction} \label{sec:intro}

At birth, stars and planets are imparted an initial mass and an initial angular momentum. The mass fundamentally determines the fate of the objects -- it sets the lifetime, radiation output, evolutionary path, and interior structure. The role of the initial angular momentum on the other hand is less obvious. Low-mass stars do not conserve angular momentum. In the first few million years of their evolution stars like the Sun shed orders of magnitude of angular momentum through interaction with circumstellar disks and magnetic winds. Once on the main sequence, stars with spectral types F to K converge to a tight spin-mass relation where rotation period increases with stellar mass, a relation set by the physics of the wind \citep{2007prpl.conf..297H,2014prpl.conf..433B}. At this point the initial rotational conditions have been erased. 

Planets, on the other hand, are expected to retain their primordial angular momentum, as long as they are not affected significantly by tidal interaction with their host star or with moons. All planets in the solar system which fulfill this condition (the gas giants plus Mars) show a clear power-law relation between angular momentum and mass, which can also be observed between rotational velocity and mass \citep{2014Natur.509...63S}. This trend has been discussed for several decades in the solar system literature. As a possible explanation, \citet{2003P&SS...51..517H} suggest that planets accrete material from the disk until their equatorial velocity reaches a set fraction of the escape velocity, at which point accretion stops. The planetary spin-mass trend is usually thought to arise in the formation process and not in further evolution \citep[e.g.][]{1993Icar..103...67D,2014prpl.conf..595R}.

It has long been known that brown dwarfs in their rotational history are more comparable to giant planets than to solar-mass stars \citep{2009AIPC.1094...61S}. Substellar objects do spin down as they age, but the rotational braking due to winds is very weak compared to stars (factor of 10000, \citet{2014prpl.conf..433B}). The braking due to the disks is also less efficient than in stars \citep{2005A&A...430.1005L,2015ApJ...809L..29S}. As a result, brown dwarfs (and, in fact, some very low mass stars, \citet{2016ApJ...821...93N}) retain fast rotation rates of $<1$\,d for gigayears. Their angular momentum, particularly at young ages, may therefore give us insights into the formation process. 

Previously we have published rotation periods for sub-stellar objects in Upper Scorpius, a region with a median age of $\sim 10$\,Myr \citep{2012ApJ...746..154P}, using lightcurves from the Kepler/K2 mission \citep{2015ApJ...809L..29S}. Here we extend that work and present rotation periods for a sample of young brown dwarfs in the molecular clouds of Taurus, a significantly younger population with expected ages of $\sim 1$\,Myr \citep{2009ApJ...704..531K,2012MNRAS.419.1271S}. Our targets are all well characterised with spectroscopy, infrared photometry, and high-resolution imaging (Sect. \ref{sec:samp}), giving us an opportunity to look for links between substellar properties and rotation. We use several independent approaches to measure rotation periods (Sect. \ref{sec:lc}). We investigate the link between rotation and the presence of disks and find evidence for disk braking (Sect. \ref{sec:disks}). The spins of brown dwarfs, once calculated forward to their final radii, falls onto the planetary $v_{eq} \propto \sqrt{M}$ relation, in clear contrast to stars, indicating that spin rates of young brown dwarfs are predominantly set by the initial conditions (Sect. \ref{sec:spinmass}). The fact that this relation is robust over five order of magnitudes in mass highlights the fundamental importance of accretion across a wide range of formation environments.

\section{The sample} \label{sec:samp}

The Kepler/K2 mission \citep{2014PASP..126..398H} observed the Taurus star forming region in campaign 13, from March 8 2017 to May 27 2017, i.e. over about 80 days. Included in the K2 target list for this campaign was a sample of 44 young very low mass sources, as part of program GO13011 (PI: A. Scholz). As the focus of this paper is on the rotation of brown dwarfs, we limit our analysis to the objects with an estimated mass in the substellar domain. To select this subsample, we obtained the near-infrared photometry from 2MASS \citep{2003yCat.2246....0C} and spectral types from \citet{2010ApJS..186..111L} and \cite{2010ApJS..186..259R}. We de-redden the J-band magnitudes, using $A_v = ((J-K) - (J-K)_{\mathrm{phot}}) / 0.1844$ with $(J-K)_{\mathrm{phot}} = 1.0$, and $A_{\lambda} = (\lambda / 1.235 \mu m)^{-1.61}$ (see \citet{2012ApJ...744....6S} for a justification of these parameters). The de-reddened J-band magnitudes were converted to absolute magnitudes $M_J$ assuming a distance of 140\,pc \citep{2009ApJ...698..242T}.

To select brown dwarfs, we require $M_J>6.0$ or spectral type Mx with $x>=6.0$. According to theoretical isochrones \citep{2015A&A...577A..42B}, these criteria imply masses below or around the substellar threshold. We find 25 objects satisfying these criteria -- this constitutes the primary sample for this paper. Their properties are listed in Table \ref{tab:table1} for reference, including H$\alpha$ equivalent widths (from the literature) and effective temperatures, calculated from the spectral type. For the latter conversion, we used the relation from \citet{2014ApJ...785..159M}, which was derived using M dwarfs in Lupus, a region with an age similar to Taurus. As shown in \citet{2014ApJ...785..159M}, this relation is generally consistent with similar conversions in the literature, and suggests a typical error of $\pm 100$\,K in $T_{eff}$.

\begin{figure}[t]
\centering
\includegraphics[width=0.9\textwidth]{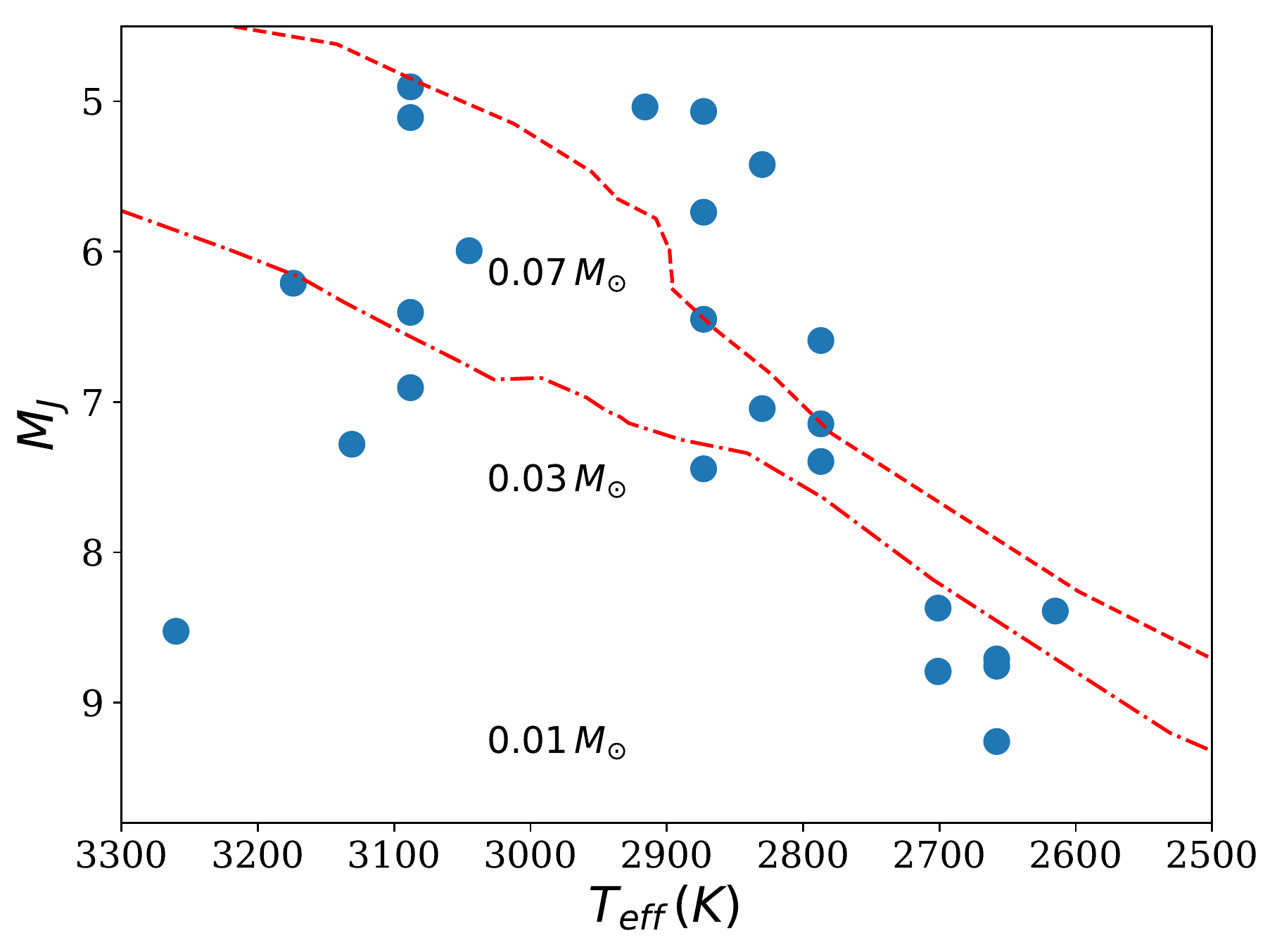}
\caption{\label{fig:hrd} Hertzsprung-Russell Diagram for the sample of young brown dwarfs in Taurus. Overplotted with red lines are the 1 (dashed) and 5\,Myr (dash-dotted) isochrones from \citet{2015A&A...577A..42B}. Approximate mass limits, calculated from $M_J$, are indicated.}
\end{figure}

In Fig. \ref{fig:hrd} we show the Hertzsprung-Russell Diagram for the sample, in comparison with theoretical isochrones for ages of 1 and 5\,Myr \citep{2015A&A...577A..42B}. Most objects fall in the area close to these two lines. Noteworthy is a datapoint far below the 5\,Myr isochrone; this is EPIC247591534 (a disk-bearing, accreting Taurus member) which appears to be too hot for its brightness or too faint for its temperature. This could be a low-mass star seen through an edge-on circumstellar disk. As far as we are aware, none of the objects in this list is known to have a companion, although the majority of them have been observed with high spatial resolution $<0.1"$ \citep{2012ApJ...757..141K,2014ApJ...788...40T}.

\begin{table}
\caption{\label{tab:table1} Basic properties of our primary sample.}
\begin{tabular}{llllllccr}
\hline
EPIC & RA      & Dec     & Name & SpT$^1$ & $T_{eff}$  & $M_J$ & H$\alpha$ EW \\
     & J2000 & J2000 &      &         & K        & mag & \AA        \\
\hline
247548866 & 04 35 41.83 & +22 34 11.5 & KPNO-Tau 8     & M5.75    & 3131 & 7.28 & 15$^3$ \\
247575958 & 04 33 09.45 & +22 46 48.7 & CFHT-BD-Tau 12 & M6       & 3088 & 6.40 & 79.65$^7$ \\
247581233 & 04 35 51.43 & +22 49 11.9 & KPNO-Tau 9     & M8.5     & 2658 & 9.26 & 0.7$^3$\\
247591534 & 04 35 57.61 & +22 53 57.4 &                & M5$^2$   & 3260 & 8.52 & 21.58$^7$\\
247600777 & 04 36 38.93 & +22 58 11.9 & CFHT-BD-Tau 3  & M7.75    & 2787 & 7.40 & 43$^3$ \\
247604448 & 04 36 10.38 & +22 59 56.0 & CFHT-BD-Tau 2  & M7.5     & 2830 & 7.04 & 7.2$^3$ \\
247630187 & 04 35 08.50 & +23 11 39.8 & CFHT-BD-Tau 11 & M6       & 3088 & 6.90 & 45.07$^7$ \\
247735103 & 04 27 45.38 & +23 57 24.3 & CFHT-BD-Tau 15 & M8.25    & 2701 & 8.79 & 18.9$^7$ \\
247739445 & 04 30 23.65 & +23 59 12.9 & CFHT-BD-Tau 16 & M8.25    & 2701 & 8.79 & 16.76$^7$\\
247748412 & 04 32 23.29 & +24 03 01.3 &                & M7.75    & 2787 & 6.59 &  \\
247791556 & 04 33 01.97 & +24 21 00.0 & MHO 8          & M6       & 3088 & 4.90 & 18/14$^4$\\
247794491 & 04 32 50.26 & +24 22 11.5 & CFHT-BD-Tau 5  & M7.5     & 2830 & 5.42 & 29.84$^7$ \\
247915927 & 04 44 27.13 & +25 12 16.4 &                & M7.25    & 2873 & 5.74 & 100$^5$\\
247950452 & 04 33 42.91 & +25 26 47.0 &                & M8.75    & 2615 & 8.39 & \\
247953586 & 04 32 03.29 & +25 28 07.8 &                & M6.25    & 3045 & 5.99 & \\
247968420 & 04 41 48.25 & +25 34 30.5 &                & M7.75    & 2787 & 7.15 & 586$^5$ \\
247991214 & 04 39 03.96 & +25 44 26.4 & CFHT-BD-Tau 6  & M7.25    & 2873 & 6.45 & 63.74$^7$ \\
248015397 & 04 41 10.78 & +25 55 11.6 & CFHT-BD-Tau 8  & M5.5     & 3174 & 6.21 & 52$^7$ &\\
248018652 & 04 30 57.18 & +25 56 39.4 & KPNO-Tau 7     & M8.25    & 2701 & 8.37 & 122$^3$ \\
248023915 & 04 38 00.84 & +25 58 57.2 & ITG 2          & M7.25    & 2873 & 5.07 & \\
248029954 & 04 39 47.48 & +26 01 40.7 & CFHT-BD-Tau 4  & M7       & 2916 & 5.04 & 79$^3$\\
248044306 & 04 30 07.24 & +26 08 20.7 & KPNO-Tau 6     & M8.5     & 2658 & 8.76 & 77.5$^3$\\
248051303 & 04 38 14.86 & +26 11 39.9 &                & M7.25    & 2873 & 7.44 & 47$^6$\\
248053986 & 04 33 52.45 & +26 12 54.8 &                & M8.5     & 2658 & 8.71 & \\
248060724 & 04 33 07.80 & +26 16 06.6 & KPNO-Tau 14    & M6       & 3088 & 5.11 & 22.1$^3$ \\
\hline
\end{tabular}
$^1$\citet{2010ApJS..186..111L}, $^2$\citet{2010ApJS..186..259R} , $^3$\citet{2005ApJ...626..498M}, $^4$\citet{2003ApJ...592..266M}, $^5$\citet{2008ApJ...681..594H}, $^6$\citet{2005ApJ...625..906M}, $^7$\citet{2006A&A...446..485G} \\
\end{table}

\section{K2 lightcurves} 
\label{sec:lc}

The basis for the lightcurve analysis were the PDCSAP and K2SFF lightcurves, both downloaded from MAST. The former are the result of the 'Pre-search Data Conditioning' module applied to 'Simple Aperture Photometry', both part of the Kepler pipeline \citep{2010SPIE.7740E..1UT,2010ApJ...713L..87J}. The K2SFF lightcurves are a high-level product tailored for K2, based on the algorithm by \citet{2014PASP..126..948V}, which corrects for the pointing-dependent nature of the fluxes from K2 and achieves higher photometric precision than the original lightcurves. For our purposes, the two sets of lightcurves yield comparable results. 

A cursory visual examination of the Taurus K2 properties shows a wide range of variability. About half of our sample has an obvious period over most of the lightcurve which can be estimated by eye, in the range of 0.7-4.4\,d. Others exhibit periodic signals in parts of the lightcurve and aperiodic variability in others. Signatures of flares or bursts (i.e. a rapid increase followed by a more gradual decline) are seen as well in some cases. A small subset shows unremarkable lightcurves without coherent structure. In this paper we are primarily interested in extracting periodic signals caused by spots on the surface, which should have an approximately sinusoidal modulation and give us the rotation periods of the brown dwarfs. To account for the complexity of the lightcurves, we search for periods using several independent techniques and check the outcome carefully by visual inspection. The results of the various approaches are summarised in Table \ref{tab:table2}.

\begin{table}
\caption{\label{tab:table2} Results of the period search. Together
with the periods from the three algorithms, we list as $N$ the number of
lightcurve segments where the same period was found by ACF and as $\Delta P$ their
standard deviation. The last column contains the adopted period.}
\begin{tabular}{lccccccc}
\hline
EPIC & $P_{ACF}$  & $N$   & $\Delta P$ & $P_{GLS}$ & $P_{GP}$ & $P_{fin}$ \\
     & (d)        &       & (d)        & (d)       & (d)      & (d)        \\
\hline
247548866 & 0.69 & 7 & 0.03 & 0.69  & $0.69\pm 0.00$ & 0.69\\
247575958 & 3.87 & 2 & 0.01 & 3.49  & $3.51\pm 0.18$ & 3.51\\ 
247591534 & 1.16 & 3 & 0.02 & --    & --             & 1.16\\ 
247600777 & 0.97 & 6 & 0.03 & 0.96  & $0.96\pm 0.01$ & 0.96\\ 
247604448 & 2.82 & 4 & 0.07 & 2.93  & $2.91\pm 0.11$ & 2.91\\ 
247630187 & 1.45 & 7 & 0.04 & 1.50  & $1.50\pm 0.01$ & 1.50\\ 
247739445 & 1.54 & 4 & 0.05 & 1.61  & $1.61\pm 0.06$ & 1.61\\ 
247748412 & 3.29 & 4 & 0.05 & 3.37  & $3.37\pm 0.02$ & 3.37\\ 
247791556 & 1.09 & 6 & 0.03 & 1.03  & $1.03\pm 0.01$ & 1.03\\
247915927 & 4.43 & 5 & 0.02 & 4.43  & $4.48\pm 0.06$ & 4.48\\ 
247950452 & 0.71 & 5 & 0.08 & 0.73  & $0.73\pm 0.01$ & 0.73\\ 
247953586 & 2.39 & 6 & 0.04 & 2.38  & $2.39\pm 0.01$ & 2.39\\ 
247968420 & 2.87 & 5 & 0.06 & 2.92  & $2.99\pm ^{0.6}_{2.4}$ & 2.9 \\ 
247991214 & 3.19 & 4 & 0.04 & --    & $3.33\pm ^{0.2}_{0.4}$ & 3.3 \\ 
248018652 & --   &   &      & --    & $1.18\pm 0.16$ & 1.18\\ 
248023915 & 2.01 & 6 & 0.01 & 0.66  & $2.01\pm ^{2.0}_{1.3}$ & 2.0$^a$\\ 
248029954 & --   &   &      & --    & $2.93\pm ^{0.9}_{2.4}$ & 2.9\\ 
248060724 & 1.85 & 5 & 0.03 & 1.86  & $1.86\pm 0.01$ & 1.86\\ 
\hline
\end{tabular}

$^a$ uncertainties in the GP unrealistically large due to convergence issues in the MCMC 
\end{table}

\subsection{Autocorrelation function}

The autocorrelation function (ACF) records the similarity of a lightcurve with itself shifted by a timelag $\delta$. As such, it is expected to peak at $\delta =0$ and should show subsidiary peaks at $\delta = N \times P$ if a periodic signal is present in the lightcurve. The ACF has been extensively used for measurements of rotation periods \citep{2013MNRAS.432.1203M,2014ApJS..211...24M} and starspot lifetimes \citep{2017MNRAS.472.1618G} in Kepler data.

We computed the ACF for the entire K2SFF lightcurve and, to corroborate the result, in segments of 1/7 of the full lightcurve. This analysis was coded in Python using routines from {\sc astropy} \citep{2013A&A...558A..33A}, {\sc numpy} \citep{2011arXiv1102.1523V} and {\sc scipy} \citep{scipy}. We accepted a period if it was recorded consistently (within 0.1$\,d$) in at least 2 of our 7 segments -- the number of segments with the same period is then a quality criterion and listed in Table \ref{tab:table2}. A least-square sine fit using the ACF period on the segments was used to confirm the ACF period. The segments in which the period was clearly present were examined by eye, only those with visible periodicity are reported in Table \ref{tab:table2}. Typical uncertainties, calculated as the standard deviation over the periods derived in individual segments, are between 0.01 and 0.08\,d. This methodology is analogous to the analysis used by \citet{2015ApJ...809L..29S} for brown dwarfs in Upper Scorpius.

\subsection{Lomb Scargle periodogram}

The generalised Lomb-Scargle (GLS) algorithm \citep{1982ApJ...263..835S,1986ApJ...302..757H,2009A&A...496..577Z} is a variation of the Discrete Fourier Transform, where a time series is decomposed into a linear combination of sinusoidal functions. We computed the GLS periodograms for the PDCSAP lightcurves using the implementation that is part of the Python module {\sc PyAstronomy}.\footnote{\url{https://github.com/sczesla/PyAstronomy}} The highest maximum in the periodogram was selected as the most likely period. The lightcurves were plotted in phase to this period and those with clearly visible periodicity are reported in Table \ref{tab:table2}. The visual examination mostly removed periods longer than 15\,d which turn out to be spurious. 

\subsection{Gaussian Process}

We also inferred rotation periods with a Gaussian Process (GP) using a methodology similar to the one presented in \citet{2018MNRAS.474.2094A}. This approach is slower than the ACF and the Lomb-Scargle methods, but like the ACF it allows for non-sinusoidal, evolving variability patterns, and in addition it enables us to evaluate the posterior distribution over the period, and thus to obtain meaningful error estimates. 

We first pre-processed the PDCSAP light curves using {\sc K2SC} \citep[][\url{https://github.com/OxES/k2sc}]{2016MNRAS.459.2408A} to remove pointing-related systematics while preserving astrophysical variability, then normalised each light curve by dividing it by its median. We then analysed the systematics-corrected light curves using a quasi periodic GP model, fitting for the period alongside the other parameters of the model. GP regression, its application to stellar light curves, and its performance for measuring stellar rotation periods in Kepler/K2 data, are discussed extensively in \citet{2018MNRAS.474.2094A} and references therein, so we give only a brief description of the procedure here. Each normalised light curve is modelled as a GP with a mean of unity and a quasi-periodic covariance function:

\begin{equation}
k(t,'t) = A \exp \left[ - \Gamma \sin^2 \left( \frac{\pi |t-t'|}{P} \right) - \left( \frac{(t-t')^2}{2 \lambda^2} \right) \right],
\end{equation}

where $k(t,t')$ is the covariance between flux measurements taken at times $t$ and $t'$, and the GP hyper parameters $A$, $\Gamma$, $P$ and $\lambda$ are the variance, periodic correlation scale, period and evolutionary time-scale of the quasi-periodic behaviour, respectively. The likelihood of the data under the GP model is then simply:
\begin{equation}
\mathrm{P}(\mathbf{y}|\mathbf{t},A,\Gamma,P,\lambda,\sigma) = \mathcal{N}(\mathbf{y}|\mathbf{1}|K),
\end{equation}
where $\mathbf{y}$ and $\mathbf{t}$ are the flux measurements and corresponding times, and the elements of the covariance matrix $K$ are given by $K_{ij} = k(t_i,t_j) + \sigma^2 \delta_{ij}$, where $\sigma^2$ is the white noise variance, and $\delta_{ij}$ is 1 if $i=j$ and 0 otherwise. $\mathcal{N}(\mathbf{x}|\mathbf{a},B)$ is the probability of $\mathbf{x}$ under a multi-variate Gaussian distribution with mean vector $\mathbf{a}$ and covariance matrix $B$. Computing the likelihood requires inverting the covariance matrix and evaluating its determinant, which can be prohibitively expensive for large datasets such as the K2 light curves. We therefore used the {\sc george} package, which implements the HODLR factorisation \citep{2015ITPAM..38..252A} to speed up this process considerably.

Initially, we set $P$ to the value corresponding to the peak of the Lomb-Scargle periodogram (see Section 3.2) and the other parameters to generic but plausible initial guesses. We adopt a broad log-normal prior (standard deviation 0.5 dex) over $P$, centred on the initial guess, and log-flat priors over the other parameters $A$, $\Gamma$, $\lambda$ and $\sigma$, which merely serve to restrict their values to physically plausible ranges. We first maximise the posterior probability (likelihood times prior) with respect to all the parameters simultaneously using the Nelder-Mead algorithm as implemented in {\sc Python}'s {\tt scipy.optimize.minimize} function. We then use an affine invariant Markov Chain Monte Carlo (MCMC) sampler, implemented in the {\sc emcee} package \citep{2013PASP..125..306F} to evaluate the multi-dimensional posterior distribution. 

We visually checked the results of the GP fitting by plotting one- and two-dimensional posterior distributions for all the parameters and by comparing samples from the posterior GP distribution to the observations. We consider that we have a reliable period detection when a) the MCMC posterior distributions for all parameters, but most importantly the period, are unimodal, b) the best-fit parameters (those that maximise the posterior) allow for genuinely periodic behaviour (in particular, the evolution time scale $\lambda$ is considerably longer than the period $P$), and c) the light curve is well described by the GP model (i.e. the posterior samples appear, subjectively, to capture the variability in the light curve adequately). Based on these criteria, a convincing period detection was achieved in 14 of the light curves in our sample, listed in Table \ref{tab:table2} and a more tentative detection for a further 8 light curves. Three of the tentative detections (247991214, 248023915, 248029954) are also supported by other period search methods, visual check, or vsini (see below) and are therefore included in Table \ref{tab:table2}.

\subsection{Adopted periods}

\begin{figure}[t]
\centering
\includegraphics[width=16cm]{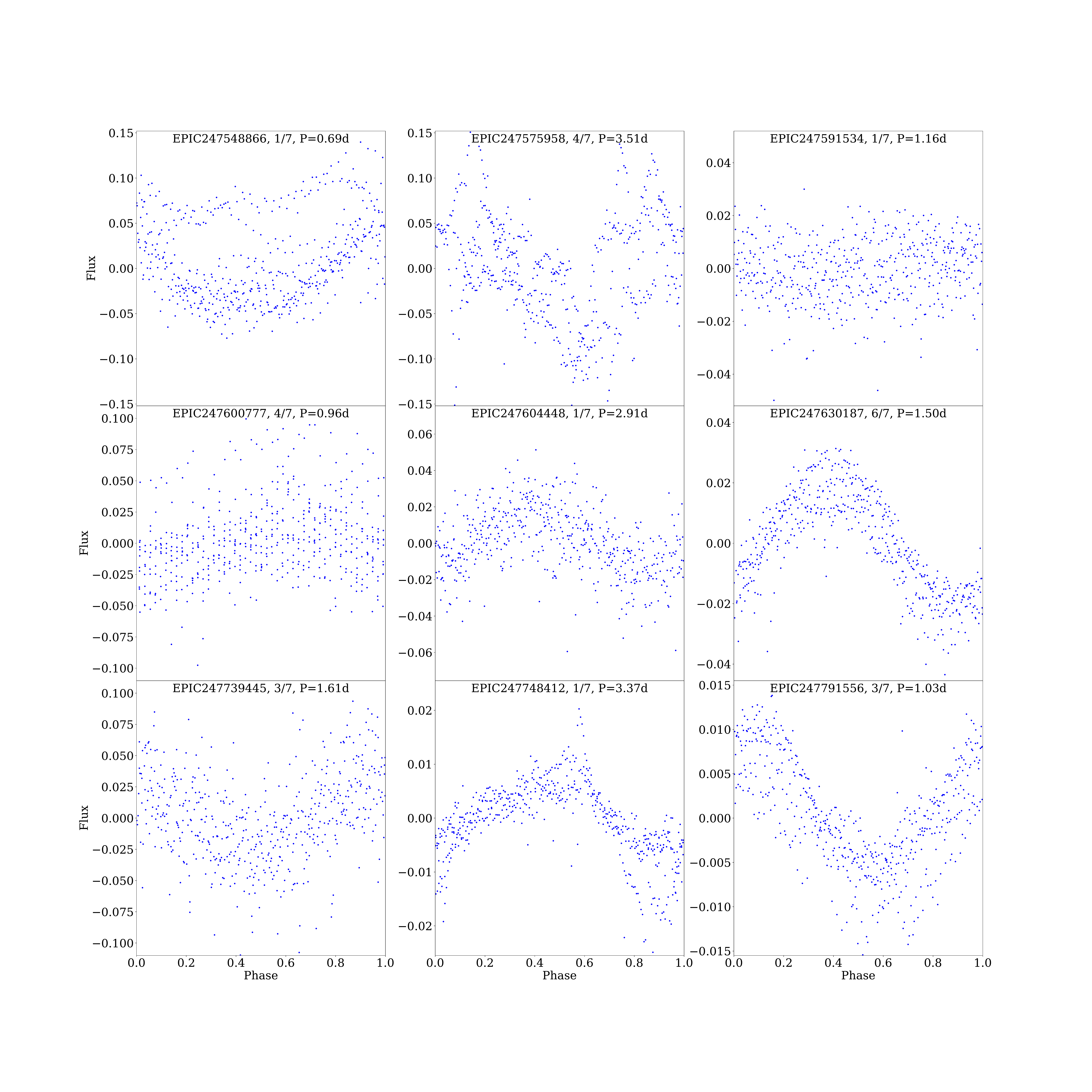}
\caption{\label{fig:phaseplot1} K2SFF lightcurves for each of the 18 objects with robust period (part 1). In each panel we show one segment (1/7th) of the full dataset. The EPIC number, segment, and adopted period is indicated. The flux average has been subtracted, and the datapoints are plotted in phase to the adopted period.}
\end{figure}

\begin{figure}[t]
\centering
\includegraphics[width=16cm]{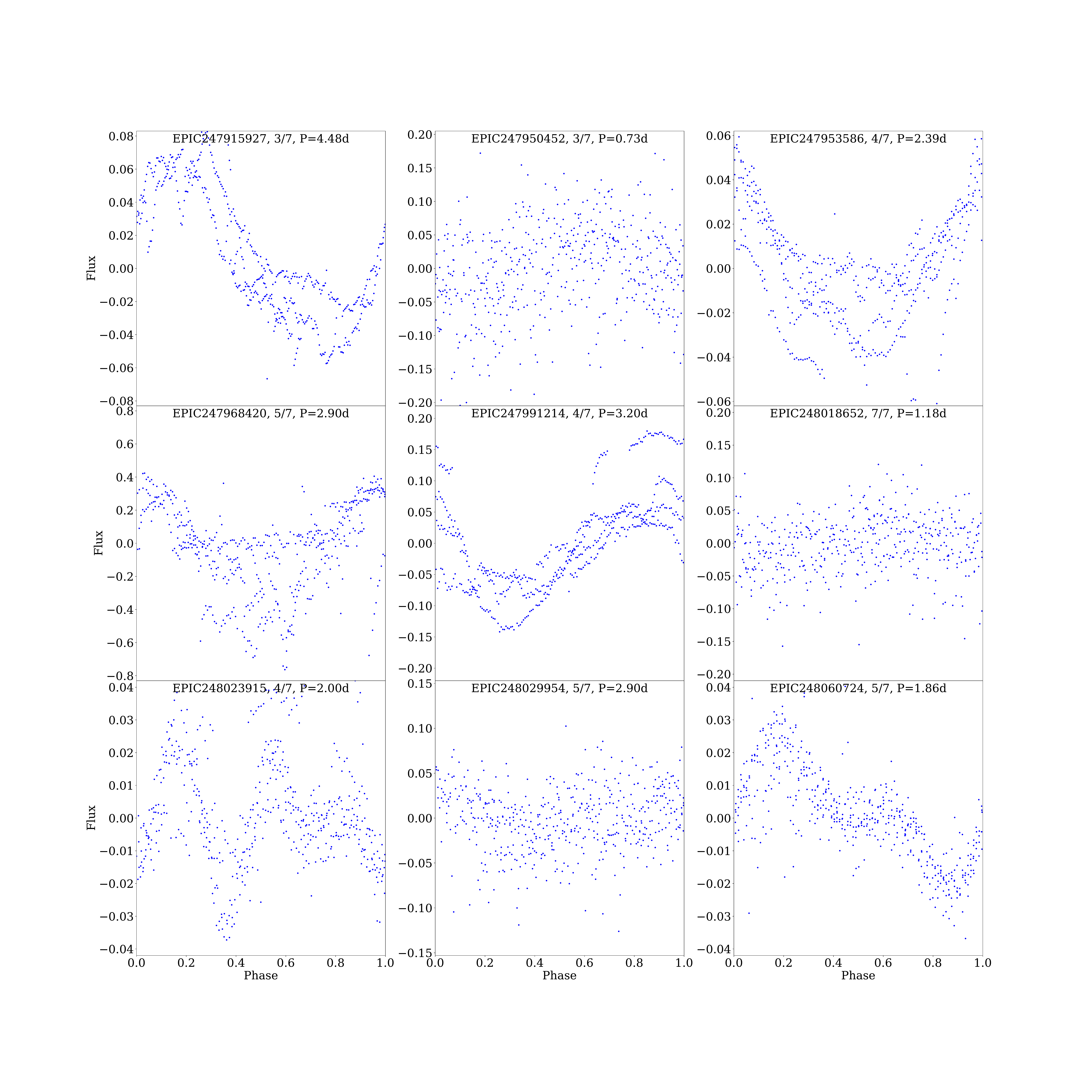}
\caption{\label{fig:phaseplot2} K2SFF lightcurves for each of the 18 objects with robust period (part 2). In each panel we show one segment (1/7th) of the full dataset. The EPIC number, segment, and adopted period is indicated. The flux average has been subtracted, and the datapoints are plotted in phase to the adopted period.}
\end{figure}

As final sample we choose the periods where we have sufficient evidence to be confident that a signal is present in the lightcurve. In all cases, the visual examination supports the conclusion that the signal is caused by modulation by spots (i.e. is mostly sinusoidal). For reference, the phased lightcurves for one segment (i.e. 1/7th of the full lightcurve) for each of the 18 adopted periods are shown in Fig. \ref{fig:phaseplot1} and \ref{fig:phaseplot2}. These plots show the periodicity clearly in most cases. In a few cases, the period is less obvious due to changes in amplitude, shape, and phase, plus additional variability. The sample of adopted periods is summarised in Table \ref{tab:table2}. Typical uncertainties in the adopted periods are $<0.1$\,d, with the exception of four cases with errors $>0.1$\,d (and thus one less significant digit in the table).

For 12 objects, all three algorithms give consistent periods within the uncertainties. For two more (248023915, 247991214) the periods from ACF and GP are consistent. For 248023915, the lightcurve shows 2-3 maxima of different height which repeat in a 2\,d cycle. This could be a signature of a complex spot distribution; the most likely rotation period is therefore 2\,d. One more (247575958) gives consistent results around 3.5\,d from GLS and GP, and a slightly larger period of 3.87\,d from ACF, only detected in 2 segments. In other segments the 3.5\,d period is more plausible. For another object, 247591534, the ACF finds a convincing period of 1.16\,d in multiple segments, which we also adopt. For two objects, 248018652 and 248029954, only the GP detects convincing periods, also supported by (secondary) peaks in the Lomb-Scargle periodogram (and, in the second case, by $v\sin{i}$, see below). This gives us a total of 18 periods. When the GP provides a clear-cut period, we adopt this value, if not we resort to the period from ACF. For the remaining 7 objects in our sample, we do not report a period.

\subsection{Comparison with rotational velocities}

For a small subset of our sample projected rotational velocities $v\sin{i}$ are available in the literature, measured from high-resolution spectroscopy by \citet{2005ApJ...626..498M}, which are a useful sanity check for our rotation periods. In Table \ref{tab:table3} we compare the derived periods with the rotational velocities and derive the implied projected radii $R\sin{i}$. For each of these objects, we also estimate the radius from the 1\,Myr isochrone by \citet{2015A&A...577A..42B}, using $T_\mathrm{eff}$ and $M_J$ from Table \ref{tab:table1}. Empirically, 1\,Myr old brown dwarfs with masses between 0.03 and 0.06$\,M_{\odot}$ are expected to have radii between 0.5 and 0.7$\,R_{\odot}$ \citep{2006Natur.440..311S}. Compared to these numbers the projected radii in Table \ref{tab:table3} look plausible, considering that $R\sin{i}$ constitutes a lower limit to the radius and that radii in this sample are expected to scatter due to the combined effects of age spread and magnetic fields \citep{2012ApJ...756...47S}. 

\begin{table}
\caption{\label{tab:table3} Periods, projected rotational velocities from \citet{2005ApJ...626..498M}, and inferred projected radii.}
\begin{tabular}{lrrll}
\hline
EPIC & period & $v\sin{i}$   &  $R\sin{i}$  & $R_{\mathrm{mod}}$\\
     & (d)    & (kms$^{-1}$) & ($R_{\odot}$) & ($R_{\odot}$) \\
\hline
247548866 & 0.69  & $45\pm3$    & $0.61 \pm 0.04$ & 0.5-1.4 \\
247600777 & 0.97  & $12\pm2$    & $0.23 \pm 0.04$ & 0.4-0.6 \\
247604448 & 2.91  & $8\pm2$     & $0.46 \pm 0.12$ & 0.5-0.6 \\
247791556 & 1.03  & $16.7\pm2$  & $0.34 \pm 0.04$ & 1.2-1.3 \\
248029954 & 2.93  & $11\pm 2$   & $0.64 \pm 0.12$ & 0.9-1.2 \\
\hline
\end{tabular}
\end{table}

\section{Rotation vs. disks} 
\label{sec:disks}

The star-disk interaction is thought to be the primary process by which low-mass stars lose angular momentum in the first few Myrs of their existence. The details of this mechanism are debated in the literature \citep{2014prpl.conf..433B}, but the observational evidence is unambiguous in several regions -- stars with disks are predominantly slow rotators, whereas those without show a wide range of rotation periods \citep[e.g.,][]{2006ApJ...646..297R,2006ApJ...648.1206J,2007ApJ...671..605C}. Here we test if the same trend can be observed among young brown dwarfs in Taurus.

\subsection{Infrared excess}

\begin{figure}[t]
\centering
\includegraphics[width=1.0\textwidth]{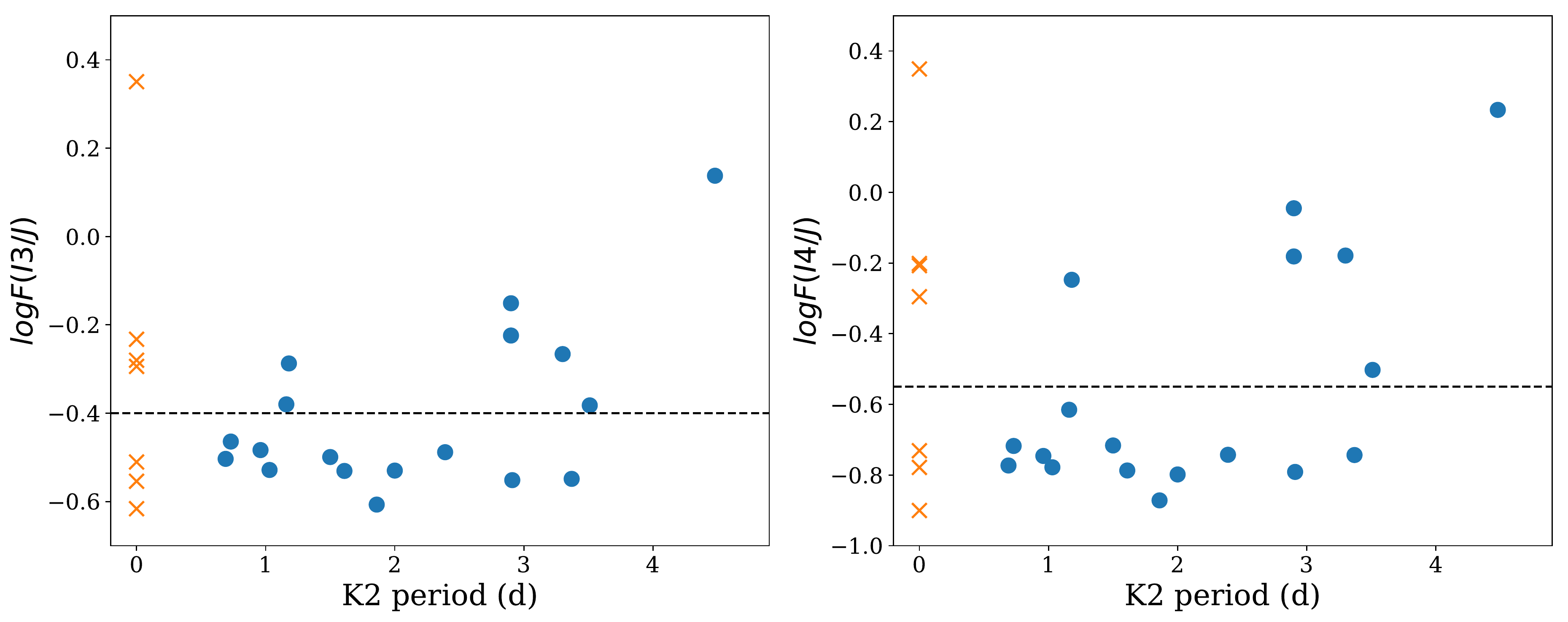}
\caption{\label{fig:ir} Infrared excess at 5.8 (left) and 8.0$\,\mu m$, measured relative to the J-band flux, vs. rotation period from K2 lightcurves. The dashed lines are the approximate delineation between objects with and without disks. Objects without measured period are plotted at $P=0.0$\,d with crosses.}
\end{figure}

We obtained Spitzer/IRAC photometry from published catalogs by \citet{2010ApJS..186..111L} or, if not listed there, by \citet{2010ApJS..186..259R}, and combine them with the 2MASS photometry to create infrared spectral energy distributions. The magnitudes were dereddened using the extinction estimated in Sect. \ref{sec:samp}, converted to flux densities $F_\nu$, and scaled by the J-band flux to obtain the IR excess. In Fig. \ref{fig:ir} we show the logarithmic IR excess at the two long IRAC wavelengths, 5.8 and 8.0$\,\mu m$, plotted against the rotation period. Objects without period measurement are plotted at $P=0.0\,h$. In terms of IR excess, the sample falls in two groups, as expected for such a young population. Objects with IR excess below the dashed lines have a spectral energy distribution that is consistent with a photosphere out to 8$\,\mu m$. Objects above that line show IR radiation exceeding the photospheric flux, an excess presumably caused by warm dust, most likely from a disk. In our sample, the fraction of objects with disks is 10 or 11 out of 25, depending on IRAC band, i.e. 40-44\%, in line with previous studies \citep[e.g.,][]{2003AJ...126...1515}. 

\subsection{Effect of disks on rotation}

In the presence of disk braking, we would expect objects with disks to be predominantly slow rotators, while objects without should show a wide range of periods, reflecting the varying time since the dissipation of the disk. The two plots seen in Fig. \ref{fig:ir} show indeed these trends. Brown dwarfs with disks have rotation periods $>3$\,d, with one or two exceptions at shorter periods. Brown dwarfs without disks have a broad range of periods up to 3.5\,d. The median period for objects with disks is 3.1\,d, vs. 1.6\,d for objects without disks. 

To put this finding on firm ground, we applied simple statistical tests. First, we carried out a two-tailed Kolmogorov-Smirnov to compare the sample of periods for objects with and without disks (using 8$\,\mu m$ excess as disk indicator). This yields a small probability of 3.0\% that the two samples are drawn from the same distribution. We verified this outcome by randomly picking 6 periods from our total sample of 18 and by checking how often we find a subsample with a median of 3.0\,d or larger, as it is found in the subsample with disks. This test tells us how likely it is that the six brown dwarfs with disks end up having on average longer periods, just by pure chance. After 10000 iterations, that likelihood is $2.8\pm0.2$\%. These tests support the conclusion that brown dwarfs with disks are preferably slow rotators.

The literature is divided on the issue of disk braking in young brown dwarfs, with all studies being affected by small samples. \citet{2004A&A...419..249S} find evidence for disk locking, but lacking sensitive mid-infrared data, they use the photometric amplitude as accretion (and thus disk) indicator, an imperfect solution. Based on spectroscopic rotational velocities and accretion indicators \citet{2005MdSAI...76...303} report that accreting brown dwarfs and very low mass stars were seen to be preferentially slow rotators compared to their non-accreting counterparts. \citet{2010A&A...515A..13R} come to a similar conclusion, but they use near-infrared photometry to detect the disks. For these cool objects, the near-infrared magnitudes are dominated by the reddened photosphere, and objects with disks do not separate sufficiently from those without. 

\citet{2010ApJS..191..389C} use for the first time mid-infrared data (as done in this paper) -- the most reliable way to probe for the presence of a disk. They see no signature of disk braking for brown dwarfs and very low mass stars in the $\sigma$\,Orionis cluster which is slightly older than Taurus, with quoted ages of 3\,Myr \citep[e.g.][]{2008AJ....135.1616S}. Thus, the most robust studies using mid-infrared data indicate disk braking at $\sim 1$\,Myr and no disk braking at 3\,Myr. That implies the disk braking timescale has to be shorter than about 3\,Myr, in line with what we found in \citet{2015ApJ...809L..29S} and significantly shorter than in stars.

For completeness, we also tested the relation between period and H$\alpha$ EW, tracing accretion. No obvious trend is visible, accretors (EW$>20$\AA, \citet{2003AJ....126.2997B}) and non-accretors have a wide range of periods. It is perhaps noteworthy that 5 of 6 objects with long periods ($>2.5$\,d) are accretors, whereas at shorter periods the ratio is 4 to 4. This may be confirmation of the weak link between the presence of disks and slow rotation found using the IR excess. 

\subsection{Comments on rotational evolution}

The sample of brown dwarf periods in Taurus gives us a new chance to examine the rotational evolution at young ages for substellar objects. The median period in our sample is 1.93\,d, ranging from 0.7 to 4.5\,d. This is very similar to the distribution of the bulk of brown dwarf periods in the (much larger, $N=139$) sample in the ONC (\citet{2009A&A...502..883R}, see their Fig. 11). Given that the population in the ONC has a similar age than Taurus, this means that different environments produce similar period distributions. 

Brown dwarfs in the older populations in $\sigma$\,Orionis \citep{2004A&A...419..249S,2010ApJS..191..389C}, $\epsilon$\,Ori \citep{2005A&A...429.1007S}, and Upper Scorpius \citep{2015ApJ...809L..29S}, ranging in age from perhaps 3 to 10\,Myr rotate somewhat faster, with median periods around 1\,d (although the different completeness of these period samples still needs to be verified). As shown in \citet{2015ApJ...809L..29S}, this difference is in line with expectations for spin-up due to contraction. In stark contrast to low-mass stars \citep{2004AJ....127.1029R,2005ApJ...633..967H}, brown dwarf periods are not 'locked' in the first few Myr of their evolution. Instead, the period evolution is consistent with angular momentum conservation, highlighting again that disk braking has to be short-lived in brown dwarfs.

\section{The spin-mass relation} 
\label{sec:spinmass}

\subsection{Period vs. mass}

It has been known for more than a decade that the typical rotation period among very low-mass stars and brown dwarfs scales with object mass \citep{2002A&A...396..513H,2005A&A...429.1007S,2017ApJ...850..134S}. Here we test this relation in our sample. In a first step, we estimate masses and radii from the absolute J-band magnitudes, by fitting 3rd order polynomials to the 1\,Myr isochrone by \citet{2015A&A...577A..42B} and applying this fit to our objects. The individual values for masses and radii carry large uncertainties in the range of $\pm 50$\%, due to a) the large empirical spread of this sample in the HR diagram (Fig. \ref{fig:hrd}) and b) the dependence on a specific evolutionary track. The latter, however, should be systematic and only in one direction, i.e. within our sample the uncertainties are likely to be smaller.

\begin{figure}[t]
\centering
\includegraphics[width=1.0\textwidth]{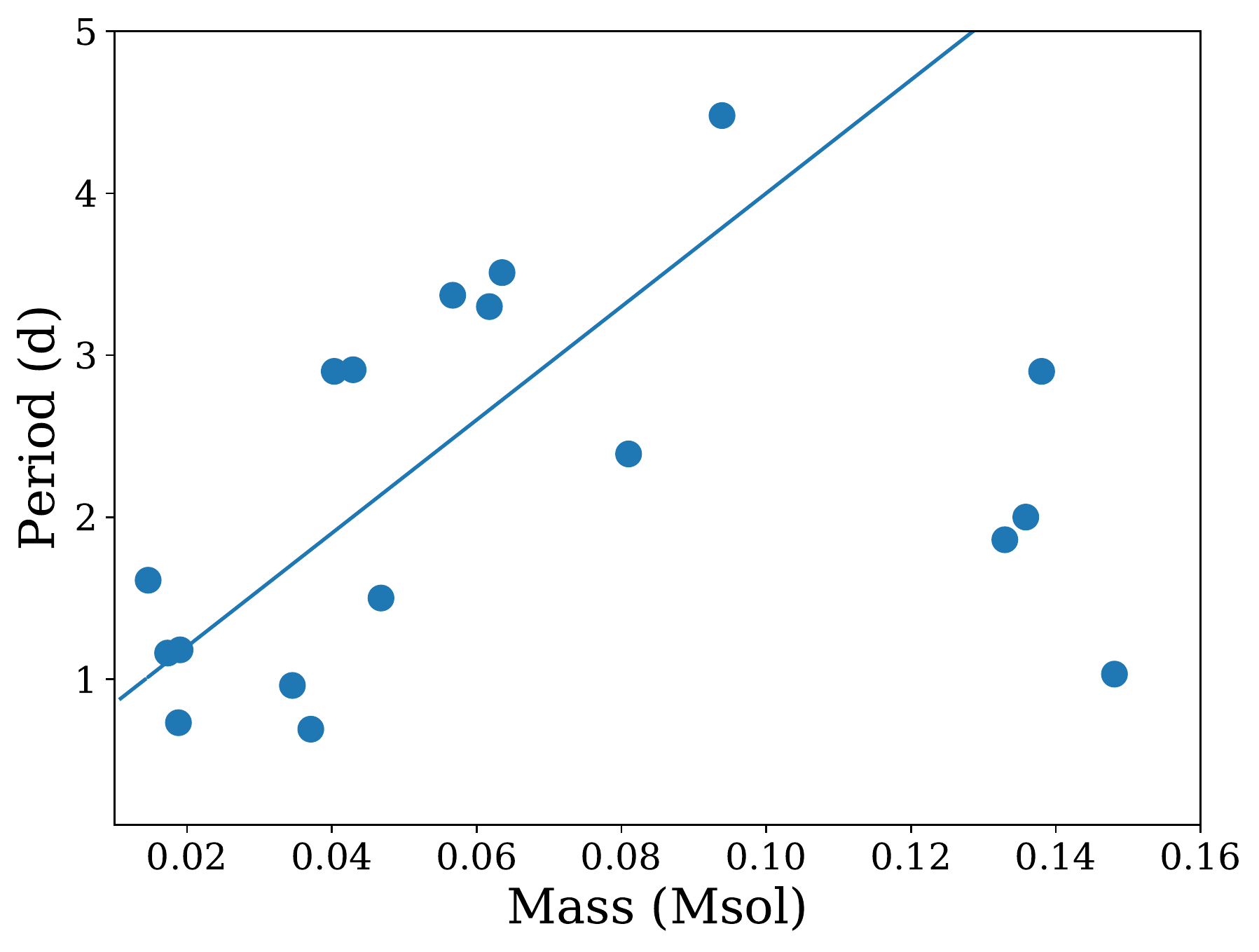}
\caption{\label{fig:permass} Rotation period plotted against estimated mass. The positive trend between mass and rotation is visible in the substellar regime.}
\end{figure}

The rotation periods scale clearly with the estimated object mass (see Fig. \ref{fig:permass}), at least in the substellar mass domain, a trend that has been established in all regions analysed so far. The fact that this relation is seen in the youngest populations -- Taurus and ONC -- points to an origin in the formation process. The mass-period relation in the substellar domain is approximately $P \sim 35 M/M_{\odot}$\,d. This is about twice as steep as in the slightly older $\sigma$\,Orionis cluster and about 8 times as steep as in the 120\,Myr old Pleiades \citep{2004A&A...421..259S}. The dropoff in the period-mass slope simply represents the spin-up due to contraction.

There are 4 objects in our sample which appear below this trend, all above the substellar boundary according to our mass estimates. These are the brightest objects in our sample, with $M_J\sim 5$. For two of them (248023915, 248029954), the spectral types (M7 and M7.25) places them robustly into the substellar domain, i.e. their J-band magnitude might be affected by excess flux (e.g., from accretion or a close companion), and hence the mass would be overestimated. For the remaining two (247791556, 248060724) the spectral type is M6 and matches what we expect from the J-band magnitudes. Given the small sample, we do not think these outliers warrant more discussion.

In contrast to the older populations in Orion \citep{2005A&A...429.1007S}, we do not find any objects rotating close to breakup speed -- the rotation period corresponding to breakup for a brown dwarf in Taurus is $\sim 0.3$\,d or below, at least a factor of 2 shorter than the fastest rotator in our sample. The breakup period develops with $R^{3/2}$, i.e. as these objects contract they will not get any closer to the breakup limit and are safely not affected by it.

\subsection{Spin-mass relation}

\begin{figure}[t]
\centering
\includegraphics[width=1.0\textwidth]{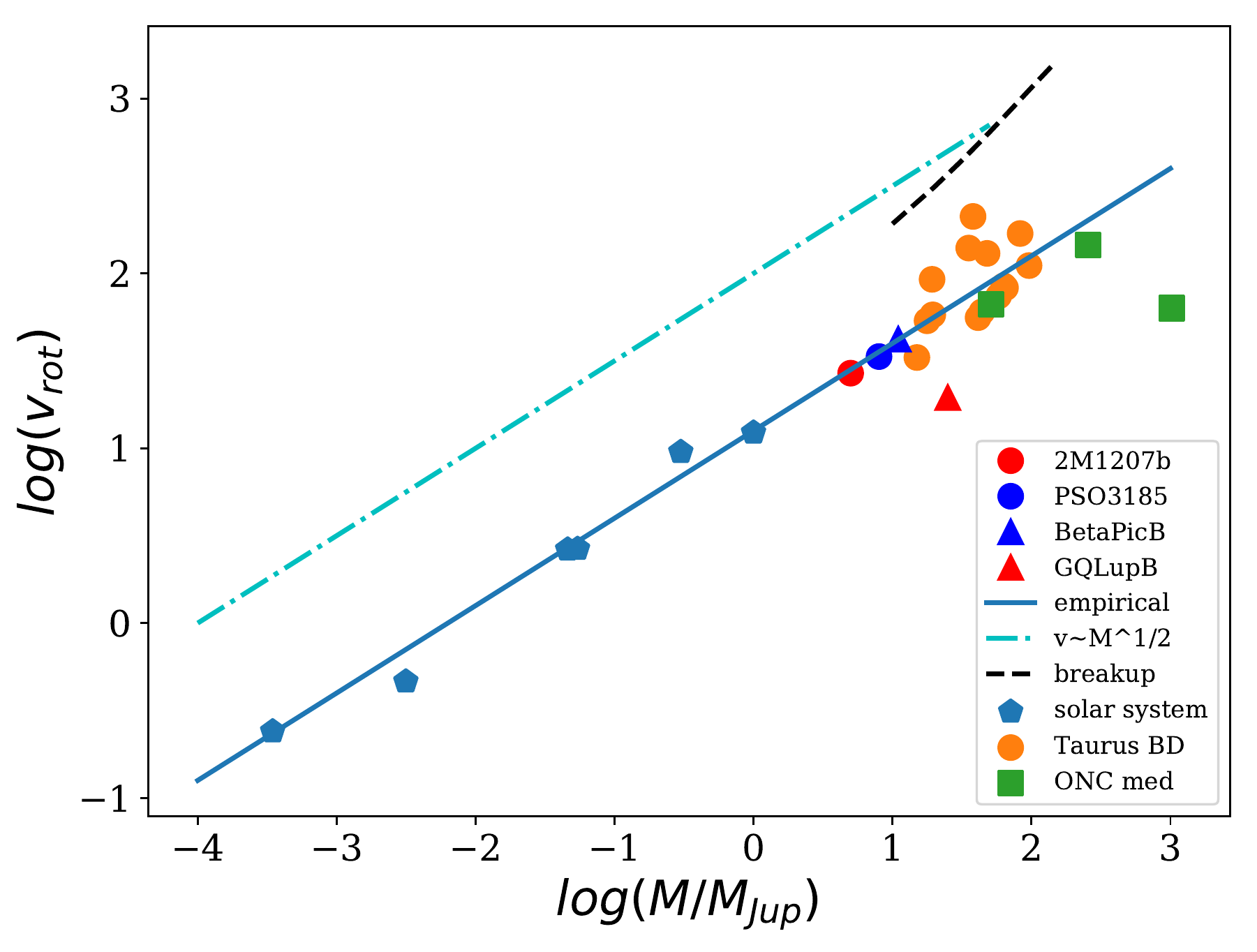}
\caption{\label{fig:spinmass} Rotational velocity vs. object mass, for solar system planets (from left to right: Mars, Earth, Uranus, Neptune, Saturn, Jupiter), the planetary mass companion 2M1207b, the free-floating planet PSO J318.5-22, the Taurus brown dwarfs, and typical brown dwarfs and stars in the ONC. For objects not in the solar system, the rotational velocity has been spun up to the age of the solar system, assuming angular momentum conservation. For reference, the breakup velocity for the age of the solar system is also overplotted.}
\end{figure}

In Fig. \ref{fig:spinmass}, we show the power law between rotational velocity and object mass for solar system planets, and extend the figure to higher masses to include brown dwarfs and stars. This is referred to as spin-mass diagram in the following. As has been discussed in the literature \citep{2003P&SS...51..517H}, the rotational velocities of solar system planets unaffected by tidal interactions (Mars, Neptune, Uranus, Saturn, Jupiter) are correlated with mass, a relation overplotted in the figure as solid line. Earth falls slightly below this trend because it has exchanged angular momentum with the Moon. Venus and Mercury are not included here because they lost angular momentum due to tidal effects. In Fig. \ref{fig:spinmass} we also show that the planetary spin-mass trend is consistent with a $v \propto \sqrt{M}$ relation. 

The first rotational information for exoplanets has also been compared with this trend. In particular, \citet{2016ApJ...818..176Z} show that the spin of the planetary mass companion 2M1207b, determined from the photometric rotation period like in this paper, fits into the planetary spin-mass trend, when accounted for the fact that this object is still contracting and will therefore spin up. The projected equatorial velocity of $\beta\,Pic$\,b is also consistent with the relation \citep{2014Natur.509...63S}. For another planetary mass companion, GQ\,Lup\,b, the measured projected equatorial velocity is significantly slower than expected from the trend, which is explained by its young age and ongoing contraction \citep{2016A&A...593A..74S}.

To put the Taurus brown dwarfs into Fig. \ref{fig:spinmass}, we calculate the rotational velocities they would have at the age of the solar system, after contracting to their final radii, assuming angular momentum conservation ('forward calculated velocities'). As final radius, we assume 0.095$\,R_{\odot}$ \citep{2015A&A...577A..42B}. For this procedure, we leave out the four objects with absolute J-band magnitudes indicating higher than substellar masses (see Fig. \ref{fig:permass}). We do the same exercise for the period of the aforementioned 2M1207b, the period of PSO J318.5-22 \citep{2017arXiv171203746B}, a young free-floating planetary mass object \citep{2015ApJ...813L..23B}, and the median period of stars in the ONC in two mass bins \citep{2009A&A...502..883R}. The projected rotational velocities for $\beta\,Pic$\,b and GQ\,Lup\,b are treated in the same way and shown as lower limits. The plot unambiguously demonstrates that the brown dwarfs fall onto the planetary spin-mass trend. Low-mass stars, on the other hand, deviate significantly from the trend, as seen in the ONC data plotted as green squares in Fig. \ref{fig:spinmass}.

The fit between forward calculated brown dwarf spin and the planetary spin-mass relation implies that, similar to planets unaffected by tidal forces, young brown dwarfs are by and large still in possession of their primordial angular momentum. This is consistent with the finding that braking by the disks is short-lived, as discussed in Sect. \ref{sec:disks}. The same cannot be said for stars, which have already lost significant angular momentum at age of 1\,Myr. As a sidenote, most field brown dwarfs end up below the relation \citep{2016A&A...593A..74S} because they, too, have lost angular momentum over the course of their evolution. While rotation braking by magnetic winds is very weak in brown dwarfs, over long timescales it does have a non-negligible effect \citep{2014prpl.conf..433B}. 

Taurus brown dwarfs with masses between 0.02 and 0.08\,$M_{\odot}$ are not expected to share a formation scenario with giant planets in the solar system and even less so with rocky planets. Predominantly they should form from core collapse followed by accretion, in a way comparable to low-mass stars \citep{2012ARA&A..50...65L}. The same probably applies to PSO J318.5-22. As isolated objects, they probably also form in a different way than the planetary-mass companions, which could have formed through disk fragmentation \citep{2005MNRAS.364L..91L}. Thus, objects from three different formation paths fit the same spin-mass relation, a powerful demonstration of the universal nature of this trend. 

There is considerable scatter around the spin-mass relation in the log-log plot. For brown dwarfs, the scatter is about 0.5 to 1.0 orders of magnitude in rotational velocity, larger than for planets. This scatter, if properly understood, can potentially reveal additional physics in the formation process, and might give insights into the differences in the formation paths of the objects discussed here. The considerable uncertainty in the mass and radius estimates (see above) also contributes to the scatter. The relation definitely deserves further empirical scrutiny beyond this paper. 

It would also be desirable to come to a more quantitative understanding of the spin-mass relation for planets and brown dwarfs, especially now that the diagram can be populated with exoplanets \citep[e.g.][]{2016ApJ...817..106B}. The escape velocity and the breakup velocity provide the right scaling of rotational velocity but due to the change in mass-radius relation across the planetary domain \citep{2017A&A...604A..83B} the continuity in the trend is puzzling. Qualitatively, the relation may originate in the physics of the accretion process and in the way the accretion is controlled and ultimately stopped. The fact that the spin-mass relation shown in Fig. \ref{fig:spinmass} seems to hold over six orders of magnitude and is obeyed by objects from several very different formation avenues points to the universal importance of accretion in the formation process of planets and brown dwarfs.

\acknowledgments
We thank the referee for a constructive report. This paper includes data collected by the K2 mission, funded by the NASA Science Mission directorate. The work was supported in part by NSERC grants to RJ. AS received funding from the UK Science and Technology Facilities Council (ST/R000824/1). SA gratefully acknowledges support from the Leverhulme Trust (RPG-2012-661) and the UK Science and Technology Facilities Council (ST/K00106X/1,ST/N000919/1). DP's work was supported by a NASA grant NNX17AI61G.

\vspace{5mm}
\facilities{Kepler/K2}

\software{ astropy \citep{2013A&A...558A..33A}, numpy \citep{2011arXiv1102.1523V}, scipy \citep{scipy}, emcee \citep{2013PASP..125..306F}, K2SC \citep{2016MNRAS.459.2408A}, PyAstronomy (\url{https://github.com/sczesla/PyAstronomy}) }

\end{document}